\documentclass[pdflatex,sn-apa]{sn-jnl}% APA Reference Style 
\usepackage{graphicx}%
\usepackage{multirow}%
\usepackage{amsmath,amssymb,amsfonts}%
\usepackage{amsthm}%
\usepackage{mathrsfs}%
\usepackage[title]{appendix}%
\usepackage{xcolor}%
\usepackage{textcomp}%
\usepackage{manyfoot}%
\usepackage{booktabs}%
\usepackage{algorithm}%
\usepackage{algorithmicx}%
\usepackage{algpseudocode}%
\usepackage{listings}%

\begin{document}

\title[Article Title]{New concept for the value function of prospect theory}

\author{\fnm{Kazuo} \sur{Sano}}\email{sano@sobin.org}

\affil{\orgname{SOBIN Institute LLC}, \orgaddress{\street{3-38-7}, \city{Keyakizaka, Kawanishi}, \postcode{ 666-0145}, \state{Hyogo}, \country{Japan}}}

\abstract{In the prospect theory, the value function is typically concave for gains and convex for losses, with losses usually having a steeper slope than gains. The neural system largely differs from the loss and gains sides. Five new studies on neurons related to this issue have examined neuronal responses to losses, gains, and reference points. This study investigates a new concept of the value function. A value function with a neuronal cusp may show variations and behavior cusps with catastrophe where a trader closes one’s position. }


\maketitle

\section{Introduction}

The prospect theory (\cite{tk1974}; \cite{kt1979}) is one of the most influential theories of human decision making (\cite{Camerer}; \cite{Goyal2021}, \cite{walter2021}; \cite{LINZMAJER2021102403}; \cite{MENGOV2022101944}; \cite{Blavatskyy2021}). According to this theory, individuals make unequal assessments of their losses and gains. Experimental studies have shown that the key behavioral assumption of expected utility theory, the so-called “independence axiom,” tend to be systematically violated in practice (\cite{vonneumann1947}). The value function is typically concave for gains and convex for losses, with losses typically having a steeper slope than gains. Decision weights are often lower than the corresponding probabilities, except in the low-probability range. Given these findings, the empirical relevance of numerous studies on the behavior of economic agents under uncertainty that uses expected utility analysis has been debated.

In economics, however, it is assumed that economic agents are typically rational. Economists have long debated the issue of time preference (\cite{strotz1955}; \cite{thaler1981}; \cite{THALER1981201}, \cite{HAIGH2005};\cite{NBERw22605}), as well as the validity of prospect theory. \cite{machina1982} showed that the basic concepts, tools, and results of his generalized expected utility analysis are independent of the independence axiom but may be derived from a much weaker assumption of smoothness of preferences over alternative probability distributions. He also showed a simple model of preferences that associates a wide body of observed behavior with risk, including the Allais paradox, may be constructed. Thus, the hypothesis that an individual maximizes a fixed preference functional defined over probability distributions may hold true for many apparently independent aspects of risk behavior. However, this may reflect only a part of human innate nature and may represent a mathematical abstract theory of human rationality in economic phenomena.

If the expected utility theory does not hold for investor’s decisions, does the neural system affect the investor trading decisions? Does the neural system respond differently to gains and losses? This study presents hypotheses to address these issues. In general, we examine the outcome of our choice and adjust subsequent choice behavior using the outcome information to choose an appropriate action. Five significant studies on neurons (\cite{KAWAI2015792}; \cite{Yamada1684}; \cite{Imaizumi2022}; \cite{Yang2022}; \cite{FERRARITONIOLO20233683}) have examined neuronal responses to loss and gain. The studies show that two different neural systems may respond to loss and gain, resulting in a value function with a cusp as a reference point.

This study addresses the reference point and cusps of value function and the following questions. When does a trader exit one’s position? Why would a trader ever close a position when losses and gains are continuously changing? The reference point is a neuronal cusp, also the value function could have other behavior cusps (\cite{RosalesRuiz1997}; \cite{Bosch2001}; \cite{Bosch2004}; \cite{smith2006}) with catastrophe (\cite{whitney1955}; \cite{Zeeman1976}).

%%%%%%%%%%%%%%%%%%%%%%%%%%%%%%%%%%%%%%%%%%

\section{Methods}

This theoretical article does not involve original data collection or analysis but rather provides a comprehensive review and synthesis of existing literature on the topic of the neural basis of the value function and decision-making. The article focuses on the prospect theory and the neural systems involved in evaluating gains and losses relative to a reference point. Five recent studies that have examined neuronal responses to gains, losses, and reference points are reviewed, and their findings are integrated into a broader framework of decision-making.

The article provides a detailed overview of the neural systems involved in decision-making, including the anterior cingulate cortex (ACC), lateral habenula (LHb), orbitofrontal cortex (OFC), dorsal striatum (DS), ventral striatum (VS), and anterior insular cortex (AIC). The article also discusses the role of these regions in the valuation of gains and losses, as well as the effects of reference points and the shape of the value function.

In addition, the article introduces a novel concept of the value function, marked by a neuronal cusp. This concept suggests a potential link between catastrophic behavior and position closure in volatile market conditions. The article highlights the need for further research to fully elucidate the implications of this concept for decision-making and financial risk management.

\section{Response of neural systems to loss and gain}

\subsection{Loss side}

The crucial role played by both lateral habenula (LHb) and anterior cingulate cortex (ACC) in monitoring negative outcomes has recently attracted attention. Kawai et al. (2015) investigated the contribution of LHb and ACC to subsequent behavioral adjustment by recording a single-unit activity from the LHb and ACC in monkeys performing a reversal learning task. The monkeys had to switch a previous choice to the alternative if the choice had repeatedly gone unrewarded in previous attempts. They found that ACC neurons stored outcome information gathered from several past trials, whereas LHb neurons detected the ongoing negative outcomes with shorter latency. Only ACC neurons signaled a behavioral shift in the next trial but not LHb neurons. Although both the LHb and the ACC represent signals associated with negative outcomes, their study findings indicated that these structures help in subsequent behavioral adjustment in different way.

Additionally, they found that ACC neurons, especially the negative outcome type, constantly accumulated the effect of negative outcome experiences on their no-reward evoked responses, which corresponded to the behavioral shift from the earlier choice to the alternative. Next, they tested the hypothesis whether the neuronal activity that accumulates negative outcome experiences triggered a behavioral shift and examined whether the no-reward-related activity predicted that the monkey would change the current choice to the alternative (next shift) or retain the current choice in the next trial (next stay). Their result is consistent with the loss side of the value function (See Figure 4 of \cite{KAWAI2015792}). From the loss perspective, a quicker detection of harm or danger is considered better suited for survival than more sensitive response to stimuli.

Although they studied monkeys and the same experiments cannot be repeated on humans, earlier studies have shown strong similarities. Such similarities are seen in human event-related potential (ERP) and functional magnetic resonance imaging (fMRI) studies (see the comprehensive review by \cite{pammi2012neuroeconomics}). As \cite{KAWAI2015792} stated, the ACC is crucial for monitoring and adjustment. Studies using human event-related potential and functional magnetic resonance imaging (fMRI) have shown that the ACC is activated when subjects receive negative feedback after an inappropriate behavioral response. However, the crucial role played by LHb in monitoring negative outcome has evoked interest. As the ACC sends projections directly to the LHb, both the structures can communicate through their reciprocal relationship. Therefore, the LHb and ACC may work together to monitor the negative outcome and alter subsequent choice behavior.

This can be understood by imagining how humans ceased living in trees and transitioned to bipedal walking. We have chosen survival strategies that develop the brain. Frogs, unlike monkeys and humans, preferred a different evolutionary strategy and aimed to improve by increasing the size of their brains. However, failure led to changing their strategy for survival by camouflaging, thus saving energy by reducing the size of their brains (\cite{Liao2022}).

\subsection{Gain side}

In their studies on brain reward circuitry, \cite{Yamada1684} and \cite{Imaizumi2022} proposed a neuronal prospect theory model. Using theoretical accuracy equivalent to that of human neuroimaging studies from gain perspective, they showed that a single-neuron activity in four core reward-related cortical and subcortical regions represents a subjective assessment of risky gambles in monkeys. Their studies focused on the central part of the orbito-frontal cortex (cOFC, area 13 M), the medial part of the OFC (mOFC, area 14 O), dorsal striatum (DS, the caudate nucleus), and ventral striatum (VS). Similar to the prospect theory framework, the attractiveness of probabilistic rewards parameterized as a product of utility and probability weighting functions is represented by the activity pattern of single neurons in monkeys passively viewing a lottery. Their study showed that the varied patterns of valuation signals were distributed in most regions of the reward circuitry and not localized. By combining these data, a network model was built to reconstruct the animals’ risk preferences and subjective probability weighting as demonstrated by their choices. Distributed neural coding thus explains how subjective valuations under risk are computed. A logistic function was used as the value function. However, the logistic function lacks a reference point, which is obvious because the researchers only focused on neuronal responses on the gain side.

Their findings about the characteristics of the gain side of the value function may be reliable. In contrast to the logistic function of the loss side, that of the gain side may seem appropriate, considering the threshold of response in the region that evokes a response to a stimulus. It is well known that fish can reduce predation risk through learning. \cite{LovnWallerius2020} showed that fish can acquire social information about a threat not only by private acquired learned experiences but also by observing the conspecifics’ response to threat and use such open information to alter future behavior through learning. This learning from the school of fish also applies to investor groups in the securities market. If you are a smart investor like Buffett, you will wait before buying a stock even if you find it very attractive. It could be a trap. It may better for survival to avoid unpleasant situations promptly and react slowly to favorable situations.

\subsection{Reference point}

\cite{Yang2022} used a token gambling task to show that monkeys altered their risk preference across wealth levels and gain/loss contexts. Consistent with the concept suggested by the prospect theory, they proposed that the AIC neurons may encode behaviorally relevant value information. AIC would represent both the subject’s current state (the reference point) and reference-dependent value signals that vary in the loss or gain context (asymmetrical value functions in loss and gain), which would together influence a subject’s risk attitude in decision making. To better understand the effect of tokens on different components, they modeled the behavior for each start token number separately. In the gain context, both monkeys were risk-seekers when the start token number was low; however, both demonstrated risk-neutral or risk-averse behavior when the start token number increased. This result is consistent with \cite{Yamada1684}, \cite{Imaizumi2022}. Moreover, \cite{Yang2022} showed, in monkey G, the utility functions (value functions) in monkey G were consistently steeper for losses than for gains. Thus, monkey G showed loss aversion behavior. However, in monkey O, the utility functions were not consistently steeper for losses than for gains, indicating that monkey O was equally sensitive to gains and losses and thus was not averse to loss. This is similar to the context where there are different types of investors.

%%%%%%%%%%%%%%%%%%%%%%%%%%%%%%%%%%%%%%%%%%
\section{Results}

\subsection{Value function}

A new concept of the value function can be easily obtained by combining these studies of neural systems’ response to losses and gains. \cite{kt1979} proposed that the value function is (1) defined on deviations from the reference point, (2) concave for gains and convex for losses, and (3) steeper for losses than for gains. However, they have also noted that “most utility functions for gains were concave, most functions for losses were convex, and only three individuals exhibited risk aversion for both gains and losses. With a single exception, utility functions were considerably steeper for losses than for gains” (\cite{kt1979}, 280). \cite{Yang2022} also found variations on the value function using two examples.  \cite{FERRARITONIOLO20233683} explored how signals from individual brain cells involved in behavior can vary. This variation could be due to the unreliability of single neurons or a more complex system where different neurons contribute different pieces of information. This concept might also be true for brain activity related to economic decisions and the rewards involved. The value of a reward is subjective and depends on both the size of the reward and the chance of getting it. To understand how the brain represents these subjective values, researchers rely on the continuity axiom. This principle is tested by presenting animals with a variety of rewards with different sizes and probabilities. If animals understand the choices and make meaningful decisions, their behavior should follow the continuity axiom.

\cite{FERRARITONIOLO20233683} explored how the brain represents subjective value in economic decision-making. Their work suggests that signals from individual neurons might be unreliable, but the brain seems to overcome this by using a "population code." This code combines information from many neurons to accurately represent subjective value, even if the individual neurons themselves have noisy or inconsistent signals.

They tested this idea in the orbitofrontal cortex (OFC), a brain region crucial for economic decisions in monkeys. They found that while individual neurons in the OFC displayed a variety of value signals, these signals often didn't directly reflect the monkey's choices. However, when they considered the activity of entire populations of neurons, a clear pattern emerged. The combined signal from the population accurately matched the monkey's choices, suggesting that the OFC uses a complex code to represent subjective value despite the limitations of individual neurons.

\begin{figure}[ht]
\centering
\includegraphics[width=8.5 cm]{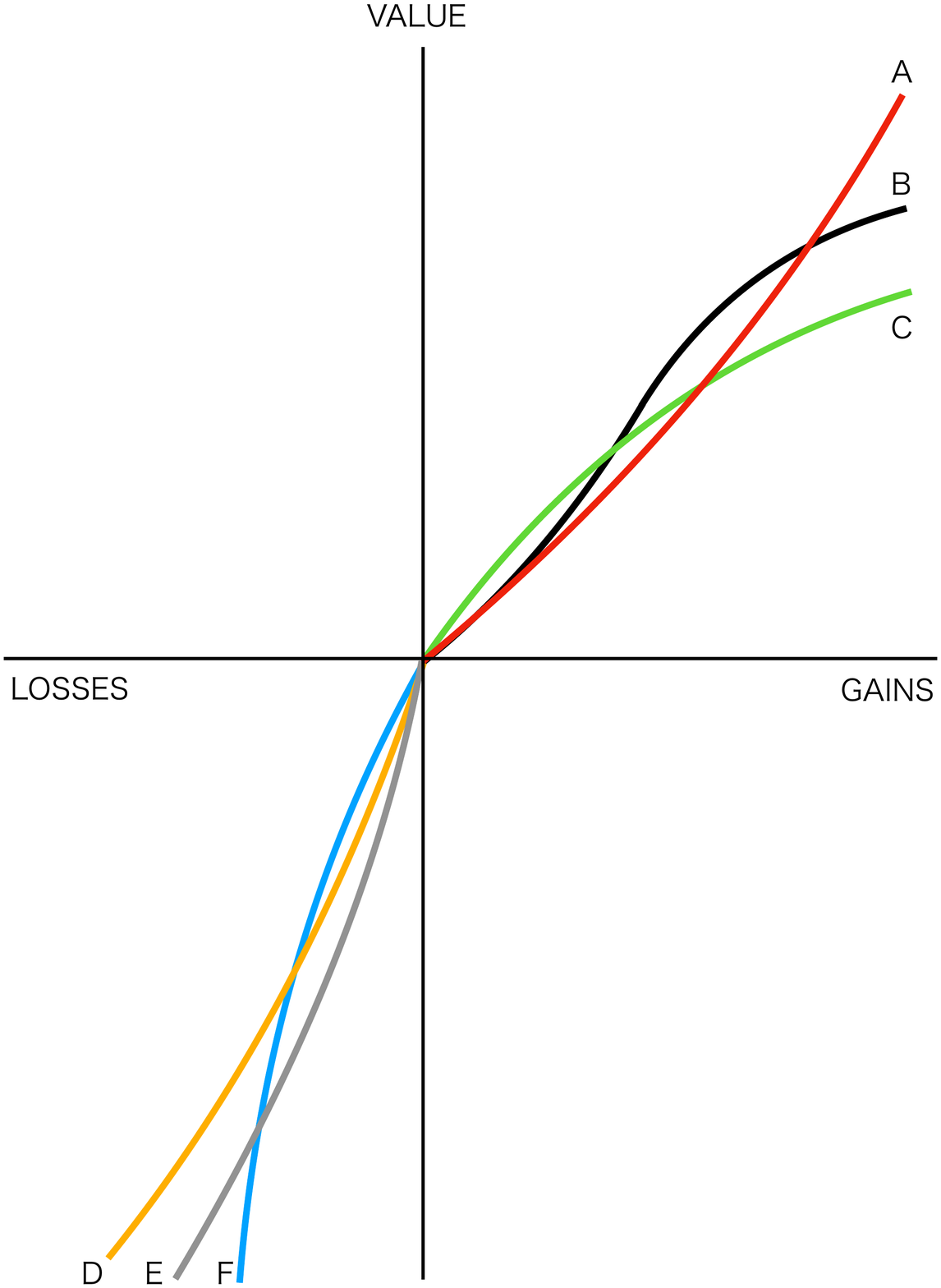}
\caption{Value functions with a reference point as a neuronal cusp. 
A: Risk loving type, B: Standard type, C: Kahneman and Tversky's original type, D: Standard type, E: Steeper type than standard, F: Reverse type of the original. We can image any other type in the area (neural systems) respond to loss and gain. This figure is essentially the same as Figure 4 b of \cite{FERRARITONIOLO20233683}, p.6  and Fig.2 b of \cite{Yang2022}, p.4.\label{fig1}}
\end{figure}  

\subsection{Behavior cusps}

In this section, we present one theoretical hypothesis regarding traders' decision-making. 
When does a trader exit one’s position? Why would a trader close a position at some point when losses and gains are continuously changing? Now, we obtain the 3D image of the value function with cusps near the decision point (Figures 2 and 3). The words “anger” and “fear” of the losing side are used in the famous dog example by \cite{Zeeman1976}. For example, when a trader of stocks exits the position, she jumps and fixes loss or gain. The value function, therefore, should have cusps corresponding to unstable psychology. It is well known investor psychology is easily affected. We can assume that these cusps are also moving. It could be a type of noise that can be synchronized and influence investors’ decisions. Indeed, when viewed individually, there are no such cusps in trading driven by mechanical algorithms. However, when viewed collectively, even trading algorithms driven by AI may involve interactions between individual trades, leading to the potential for instantaneous synchronization that could be modeled as a mathematical catastrophe.

\begin{figure}[ht]
\centering
\includegraphics[width=13 cm]{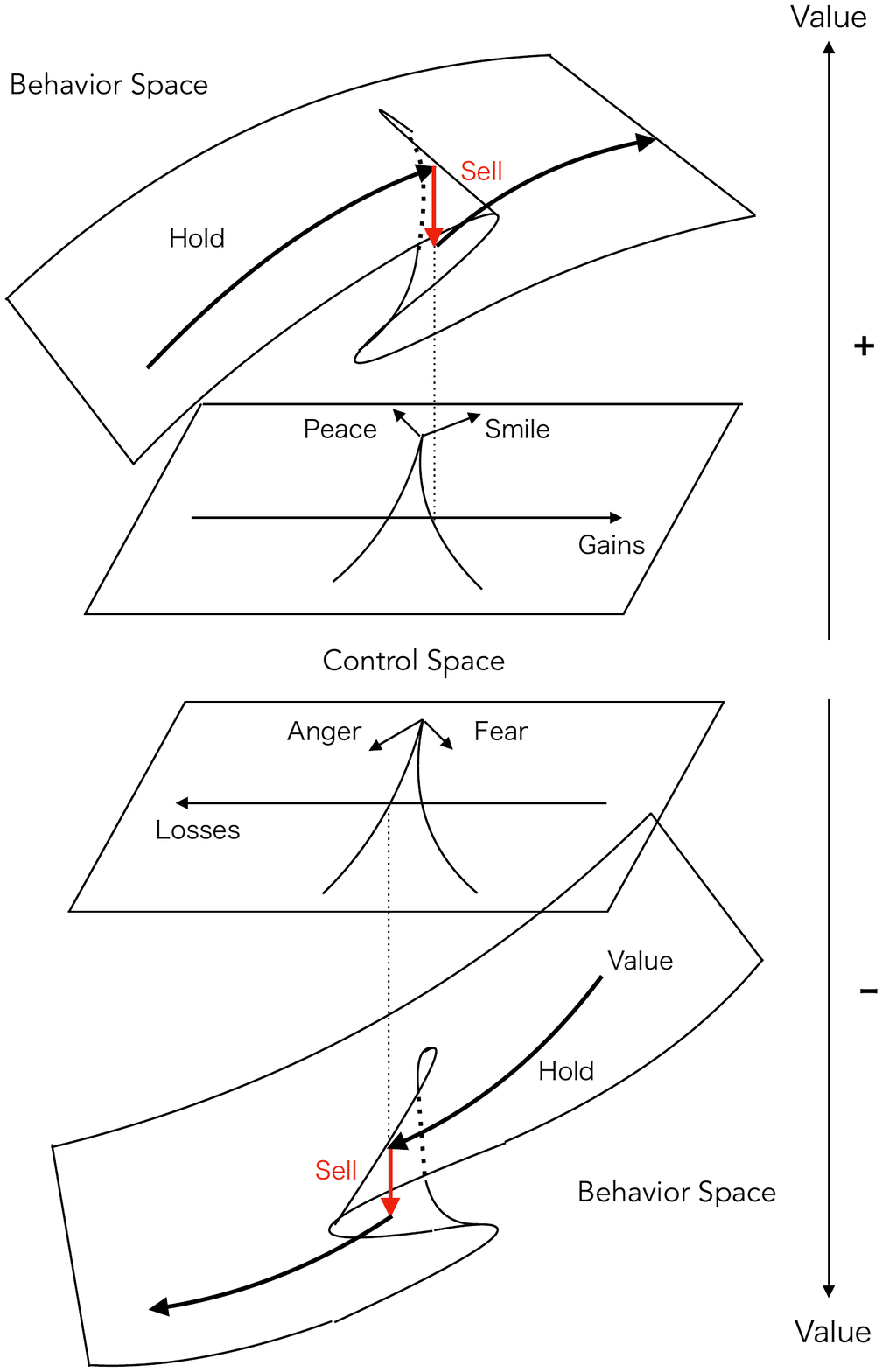}
\caption{Value function with cusps near the decision point: control space and behavior space. \label{fig2}}
\end{figure}   

\begin{figure}[ht]
\centering
\includegraphics[width=8 cm]{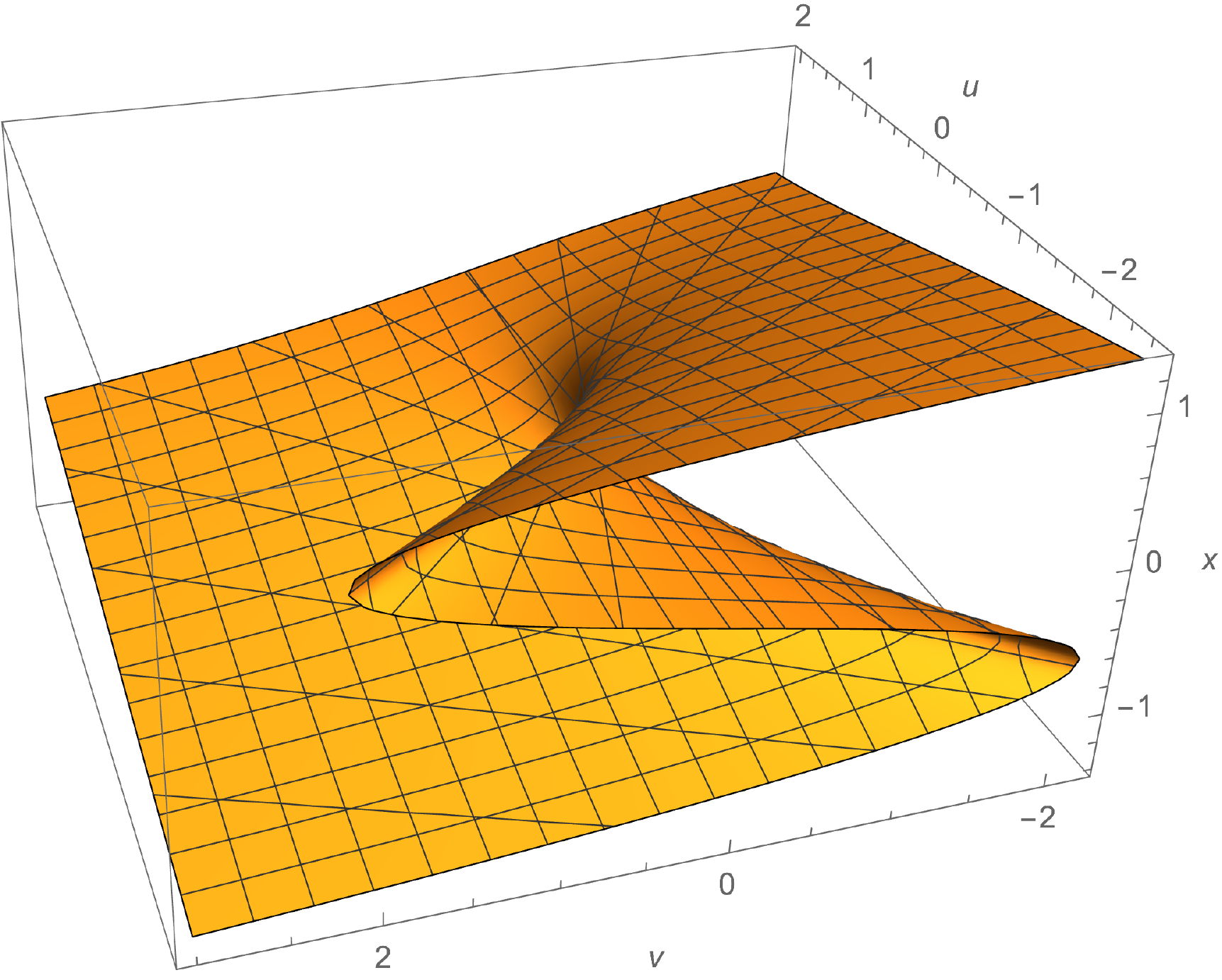}
\caption{Sample image of the cusp catastrophe: $f(x,u,v)=x^4+ux^2+vx$, by Eric W. Weisstein, 
August 3, 2021. \url{http://mathworld.wolfram.com/notebooks/DifferentialGeometry/CuspCatastrophe.nb} }  \label{fig3}
\end{figure}

%%%%%%%%%%%%%%%%%%%%%%%%%%%%%%%%%%%%%%%%%%
\section{Discussion}

This behavioral cusp remains nothing more than a theoretical hypothesis. Another issue must be resolved to accept this new image of the value function with cusps because the words “anger” or “fear” must be used to obtain the cusp representation. Since the 19th century, there has been disagreement over whether “emotions” are the cause or consequence of their associated behaviors. \cite{Anderson2014} argued that emotional behaviors are a class of behaviors that express internal emotion states. These emotional states exhibit certain general functional and adaptive properties that are true for any human emotions, such as fear or anger, and across phylogeny. \cite{PAUL2018202} stated that defining emotion in the human context is complicated, and the difficulty is magnified in animals that cannot express themselves. In these cases, notions of the internal or consciously experienced states are inaccessible. Thus, it would be premature to infer the effect of emotions from monkeys’ neuronal responses that seem apparent to us.

%%%%%%%%%%%%%%%%%%%%%%%%%%%%%%%%%%%%%%%%%%
\section{Conclusions}

The reference point, which is the value function, could be a cusp as a behavior change (\cite{RosalesRuiz1997}; \cite{Bosch2001}; \cite{Bosch2004}; \cite{smith2006}) with mathematical catastrophe (\cite{whitney1955}; \cite{Zeeman1976}). This is because the neural system substantially differs on the sides of losses and gains. We monitor the outcome of our choice and alter subsequent choice behavior using the outcome information to choose an appropriate action. \cite{KAWAI2015792}; \cite{Yamada1684}; \cite{Imaizumi2022}; \cite{Yang2022}; \cite{FERRARITONIOLO20233683} have examined the neuronal responses to losses, gains, and reference points. Thus, two neural systems may respond differently to losses and gains, resulting in a value function with a reference point as a neuronal cusp.

In economics, we assume the rationality of economic agents. For example, theoretical economics is irrelevant without rationality such as utility maximization and profit maximization. Economists have long debated issue of time preference (\cite{strotz1955}; \cite{thaler1981}; \cite{HAIGH2005}) and the validity of prospect theory. \cite{machina1982}’s generalized expected utility analysis may be used to construct a simple model of preferences, which links numerous observed behaviors toward risk. However, it is a mathematical abstract theory of human rationality in economic phenomena and represents only one side of our innate nature.

\cite{smith1759, smith1776}, considered the father of economics, believed that the pursuit of self-interest should be tempered by a “fellow feeling.” However, today’s economics focuses exclusively on the rational behavior of egocentric economic agents and does not consider of the existence and effect of empathy. 

Furthermore, \cite{smith1759}'s observation is of paramount importance when considering the value function.
\begin{quote}
    Though Nature, therefore, exhorts mankind to acts of beneficence, by the pleasing consciousness of deserved reward, she has not thought it necessary to guard and enforce the practice of it by the terrors of merited punishment in case it should be neglected. It is the ornament which embellishes, not the foundation which supports the building, and which it was, therefore, sufficient to recommend, but by no means necessary to impose. Justice, on the contrary, is the main pillar that upholds the whole edifice. If it is removed, the great, the immense fabric of human society, that fabric which to raise and support seems in this world, if I may say so, to have been the peculiar and darling care of Nature, must in a moment crumble into atoms. (SEC.II. Of Justice and Beneficence, CH.III. Of the Utility of this Constitution of Nature)
\end{quote}

According to Smith, a just society, where wrongdoing is met with punishment, is the cornerstone of social order. Punishment for wrongdoing carries more weight than simply rewarding good deeds. This aligns with the concept of a value function, where losses have a greater impact than gains. If the value function operated differently, promoting good deeds would be the foundation of justice, not punishing bad ones. Ultimately, it's our inherent sense of fairness, our empathy for justice, that forms the foundation of a stable social order.

Human empathy, the ability to share and understand the emotions of others, is another cornerstone of social interaction. Recent research has delved into the mechanisms underlying this remarkable phenomenon, uncovering the intricate interplay of shared emotions, imitation, and learned perceptual-motor links.

Imitation, another pillar of empathy, manifests as the automatic mirroring of observed behaviors. Witnessing someone's actions triggers the activation of corresponding motor representations in our own brains, as if we were preparing to perform those actions ourselves. These internally generated motor representations then guide our own actions, leading to the imitation of observed behaviors.

Recent studies on empathetic systems support Smith’s idea (\cite{Waal_2002}; \cite{BB08111221};\cite{Yamamoto2016}), which include noise synchronization (\cite{app12147019}). Maybe it is time to revisit Smith’s concepts and rethink economics in terms of justice and empathy. This perspective is lacking in modern microeconomic theory and could introduce a new field to economics. Further consideration is needed on the issue of justice and empathy as well as the value function of prospect theory.

\section*{Declarations}
\subsubsection*{Funding} Not applicable.
\subsubsection*{Ethical Approval}  Not applicable.
\subsubsection*{Disclosure statement} The author declares no conflict of interest.
\subsubsection*{Data availability} Data sharing is not applicable to this article as no datasets were generated or analysed during the current study.

%\newpage

%\bibliographystyle{unsrtnat}
\bibliography{references}  %%% Uncomment this line and comment out the ``thebibliography'' section below to use the external .bib file (using bibtex) .





% \end{thebibliography}

\end{document}